\documentclass[twoside,twocolumn,english,aps,prl,longbibliography,floatfix]{revtex4-1}
\usepackage{color}
\usepackage{textcomp}
\usepackage{amsmath}
\usepackage{amssymb}
\usepackage{graphicx}
\usepackage{natbib}
\usepackage{esint}
\usepackage[normalem]{ulem}
\usepackage{subfloat}
\usepackage[caption=false,singlelinecheck=off]{subfig}
\makeatletter
\@ifundefined{textcolor}{}
{%
 \definecolor{BLACK}{gray}{0}
 \definecolor{WHITE}{gray}{1}
 \definecolor{RED}{rgb}{1,0,0}
 \definecolor{GREEN}{rgb}{0,1,0}
 \definecolor{BLUE}{rgb}{0,0,1}
 \definecolor{CYAN}{cmyk}{1,0,0,0}
 \definecolor{MAGENTA}{cmyk}{0,1,0,0}
 \definecolor{YELLOW}{cmyk}{0,0,1,0}
}
\usepackage{times}
\usepackage[colorlinks=true,linkcolor=blue,urlcolor=blue,citecolor=blue]{hyperref} 
\hypersetup{colorlinks=true}
\usepackage[all]{hypcap} 

\begin{document}
\title{Distinguishing between Topological and Quasi Majorana Zero Modes with a Dissipative Resonant Level}
\author{Gu Zhang}
\email{gu.zhang@kit.edu}
\author{Christian Sp\r{a}nsl\"{a}tt}
\email{christian.spanslatt@kit.edu}
\affiliation{Institute for Quantum Materials and Technologies, 76021 Karlsruhe, Germany}
\affiliation{Institut f\"{u}r Theorie der Kondensierten Materie, Karlsruhe Institute of Technology, 76128 Karlsruhe, Germany}
\date{\today}
\begin{abstract}
Distinguishing between topological Majorana zero modes and quasi Majorana modes -- trivial low energy Andreev bound states -- is an important step towards realizing topological hybrid nanowire devices. We propose that this distinction can be made by connecting a hybrid nanowire to a dissipative resonant level realized by a quantum dot weakly coupled to high resistance leads. We show that the qualitative temperature scaling of the
width of the resonant level zero bias conductance peak provides a signature unique to true Majorana zero modes. The degeneracy induced by a true Majorana zero mode is further shown to stabilize this conductance peak against particle-hole asymmetry.
\end{abstract}
\maketitle
Majorana zero modes (MZMs) -- self-conjugate and gapless states confined to boundaries or defects in topological superconductors (SCs) -- have attracted enormous attention in the last two decades~\cite{Alicea2012,Beenakker2015,Elliot2015,Aguado2017}.
On top of their non-Abelian statistics and close connection to quantum computation~\cite{Ivanov2001,Nayak2008}, MZMs are expected to influence a wide variety of physical phenomena such as spin-selective Andreev reflection~\cite{He2014}, exotic single-~\cite{Kitaev2001,Fu2008,Spanslatt2015PiJunc,Spanslatt2018} and multi-terminal~\cite{Spanslatt2017,Meyer2019} Josephson effects, and non-local correlations~\cite{Fu2010}. MZMs may also arise in non-topological systems, such as boundary impurity problems, where they manifest non-Fermi liquid behaviour due to interactions and frustration~\cite{EmeryKivelsonPRB92,WongAffleck94,Mebrahtu12}.

A well-studied system proposed to host MZMs is a strong spin-orbit coupled semi-conducting nanowire~\cite{Lutchyn2010,Oreg2010}. In proximity to an $s$-wave SC, and above a critical external magnetic field, the wire is expected to enter the topological regime: effectively a spinless $p$-wave SC with exponentially localized edge MZMs~\cite{Kitaev2001}. Such an isolated edge MZM has been predicted to generate a topologically protected $2e^2/h$ zero bias conductance peak (ZBCP) due to perfect Andreev reflection~\cite{Law2009,Flensberg2010,Liu2012,Fidkowski2012}, a target of intense experimental work \cite{Deng2012,Mourik2012,HaoZhangNature18,Prada2019}.

An increasing amount of work indicates however that pairs of low energy Andreev bound states, so-called pseudo- or quasi-MZMs (qMZMs), can be generated quite generally in the trivial regime, for wide ranges of parameters~\cite{Prada2012,BrouwerPRB12,Stanescu2013,SanJose2016,Cayao2015,Liu2017,Fleckenstein2018,Moore2018,vuik2018reproducing,Stanescu2019,Prada2019NH,Awoga2019}. For instance, a smooth wire confinement potential can generate qMZMs with a coupling energy exponentially small in the inverse potential slope~\cite{BrouwerPRB12}, thereby making them difficult to distinguish from true, topologically induced MZMs. These results suggest that ZBCPs close to $2e^2/h$ could potentially come from qMZMs and can therefore not be uniquely attributed to topological superconductivity. It has however been pointed out that qMZMs can, under certain circumstances, mimic also other behaviours of MZMs, including non-Abelian statistics (see e.g. Ref.~\onlinecite{vuik2018reproducing}).
\begin{figure}[t]
\captionsetup[subfigure]{position=top,justification=raggedright}
\subfloat[]{
\includegraphics[width=0.7\columnwidth]{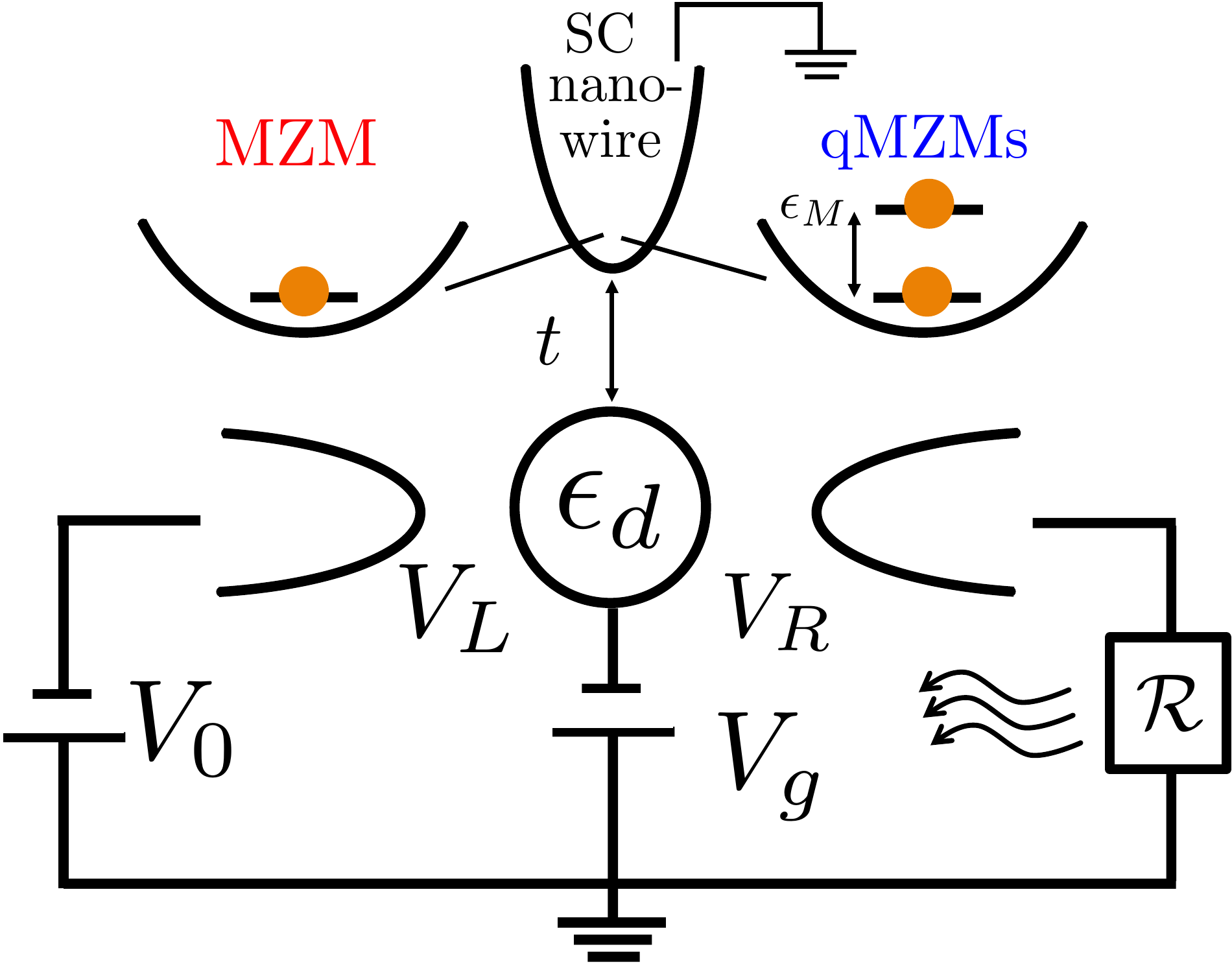}
\label{fig:Setup}}
\\[0.0cm]
\subfloat[]{
\includegraphics[width =0.7\columnwidth]{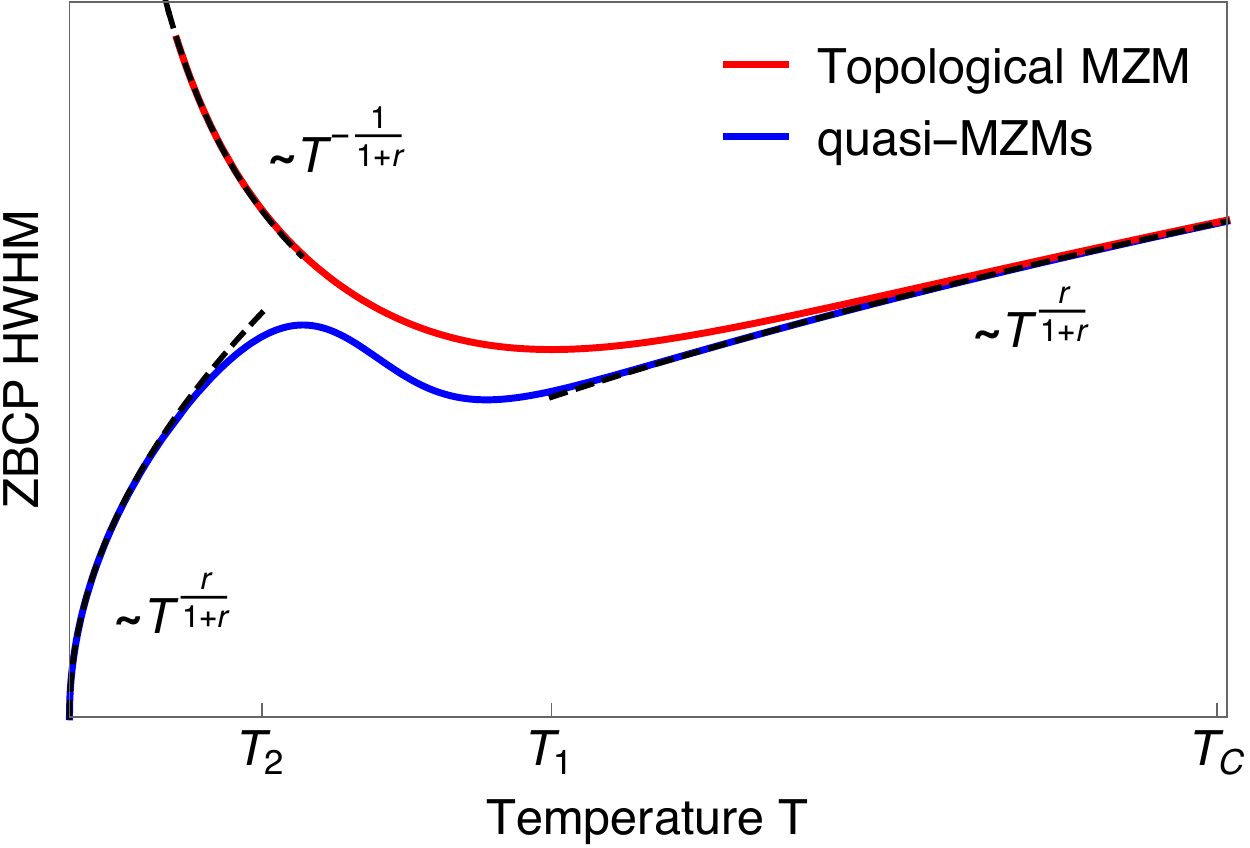}
\label{fig:Scalings}}
\caption{(a) Schematics of a hybrid nanowire coupled to a dissipative resonant level model realized by a quantum dot and high resistance leads. The lead-dot couplings are $V_{L,R}$ and the dot energy $\epsilon_d$ is tunable through the backgate voltage $V_g$. Tunneling events through the dot induce dissipation through the Ohmic impedance $\mathcal{R}$, parametrized by $r\equiv\mathcal{R}e^2/h$. The nanowire couples either a single topological MZM or a qMZM pair to the dot. (b) The half width at half maximum (HWHM) of the zero bias conductance ($dI/dV|_{V_0=0}$) peak as a function of the temperature $T$ in the strong coupling limit (i.e., below the characteristic temperature $T_c$). MZM and qMZMs induce disparate HWHM scalings with decreasing $T$. The two temperature scales $T_{1} \propto \max(t_{1}, t_{1}')$ and $T_{2} \propto \max(t_{2}, t_{2}')<T_{1}$ [see Eq.~\eqref{eq:HamiltonianFull}] mark respectively the scale where the dot-wire coupling is revealed and where a single MZM and a pair of qMZM are distinguishable.}
\end{figure}

In this Paper, we show that topological MZMs and qMZMs can be distinguished by coupling the edge of a SC nanowire to a frustrated dissipative resonant level (DRL), realized by a single level quantum dot weakly coupled to two dissipative leads (see Fig.~\ref{fig:Setup}).
MZMs in dissipative environments have been studied before~\cite{Liu2013,Liu2019,Zhang2019}, and we show here that coupling the DRL to a topological MZM or two qMZMs generate drastically different experimental features in terms of (i) the position of the \textit{low-temperature ZBCP through the DRL} and (ii) the low temperature scaling of this ZBCP width (half width at half maximum) (see Fig.~\ref{fig:Scalings}). Coupling the dot to a MZM, the ZBCP is stable even off-resonance, with its width increasing as $\sim T^{-1/(1+r)}$ with decreasing temperature. This widening saturates when the detuning becomes comparable to energy scales on the order of the SC gap or the dot level spacing. 
In stark contrast, coupling the dot to a qMZM pair, the ZBCP instead shifts away from resonance and upon lowering the temperature its width decreases as $\sim T^{r/(1 + r)}$, similarly to that of an isolated DRL model \cite{Mebrahtu12,Mebrahtu13}. These distinct features thus provide evident experimental signatures for singling out topologically induced MZMs in hybrid nanowire devices.

The basic physical mechanism underlying these disparate features is that only the true MZM provides a necessary degeneracy otherwise only obtainable by fine-tuning the dot to the particle-hole symmetric point~\cite{Mebrahtu12,Mebrahtu13}. This degeneracy uniquely stabilizes the DRL strong coupling (renormalization group, RG) fixed point. The relaxation of the particle-hole symmetry requirement is a manifestation of the topological Kondo effect, in which MZMs provide a ``non-local'' degeneracy induced by topology~\cite{Beri2012,Beri-MajoranaKleinPRL13}.
Our work thus provides an example where topology replaces an otherwise required symmetry in a boundary system.

\textit{Model.--} 
The quantum dot and dissipative leads (with Ohmic dissipation $\mathcal{R}$) realizing the the DRL is described by the Hamiltonian $H_{\text{DRL}} = H_{\text{leads}} + H_{\text{dot}} + H_{\text{T}}$, where
\begin{equation}
H_{\rm leads} = \sum_k \epsilon_k\big(c^\dag_{L,k}c^{}_{L,k}+c^\dag_{R,k}c^{}_{R,k}\big),
\label{eq:hleads}
\end{equation}
in which $c^{\dagger}_{\alpha,k}$ is the creation operator of a free electron with momentum $k$ and energy $\epsilon_k$ in lead $\alpha=L,R$. The dot Hamiltonian $H_{\rm dot} = \epsilon_d d^{\dag}_{}d^{}_{}$ includes only a single energy level $\epsilon_d$, defined with respect to the lead Fermi levels and is tunable through a back gate voltage $V_g$ (see Fig.~\ref{fig:Setup}).
Finally, the point tunneling Hamiltonian
\begin{equation}
H_{\rm T} = \sum_{k}\big(V_L c^\dag_{L,k}d e^{-i\varphi/2}+ V_R c^\dag_{R,k}d e^{i\varphi/2}+h.c. \big)
\label{eq:ht}
\end{equation}
describes dot-lead tunneling, with strengths $V_L$ and $V_R$. The phase fluctuation operator $\varphi$ is conjugate to $N_L - N_R$, where $N_{L,R}$ are lead number operators.
The operator $\exp(\pm i \varphi/2)$ thus describes charge tunneling and the associated charge relaxation that couples to the dissipative bath~\cite{IngoldNazarov92, LiuRLdissipPRB14}.
For an Ohmic dissipation with a linear spectral density \cite{LeggettRMP87}, $\varphi$ acquires effective dynamics (in time $\tau$) after integrating out the dissipative modes \cite{IngoldNazarov92}
\begin{equation}
\big\langle e^{i\varphi(\tau)} e^{-i\varphi(0)} \big\rangle \propto \frac{1}{\tau^{2r}},
\label{eq:dissipation_correlation}
\end{equation}
where $r = \mathcal{R}/\mathcal{R}_Q$ is the dimensionless dissipation in units of $\mathcal{R}_Q \equiv h/e^2$.
Noticeably, with generic parameters, the low temperature conductance $G$ of the DRL model is greatly suppressed: the so-called dynamical Coulomb blockade effect~\cite{IngoldNazarov92,NazarovBlanterBook, DevoretEsteveUrbina95}, goverened by a weak coupling fixed point. However, for a fine-tuned parameter choice $V_L = V_R = V$ and $\epsilon_d = 0$, the system flows instead towards a non-Fermi liquid strong coupling fixed point characterized by a perfect ZBCP $G=e^2/h$~\cite{Mebrahtu12, Mebrahtu13, LiuRLdissipPRB14} and a finite residue entropy \cite{WongAffleck94,Huaixiu14_1-G}.
Similarly to the two-channel Kondo model~\cite{EmeryKivelsonPRB92}, this critical point occurs due to the frustration between the two equally coupled leads as well as the symmetry induced dot degeneracy~\cite{2CK}. The sensitivity of the low temperature conductance around this point against symmetry breaking is thereby a useful platform to investigate the influence of topological MZMs.

We next introduce a low energy effective model for the SC nanowire, which provides either a single true MZM, $\gamma_1$, or a pair of weakly coupled qMZMs, $i\epsilon_M\gamma_1 \gamma_2$ (see Fig.\,\ref{fig:Setup}).
For the qMZM scenario, $\gamma_1$ and $\gamma_2$ couple to dot Majorana operators with the strengths $t_1$, $t_1'$ and $t_2$, $t_2'$, respectively~\cite{SuppMat}.
The MZM scenario then corresponds to $t_2 = t_2' = \epsilon_M = 0$.
Due to the external magnetic field, the system is effectively spinless and one of the qMZMs couples more strongly to the dot level~\cite{vuik2018reproducing}. Without loss of generality we therefore assume $t_2, t_2' \ll t_1, t_1'$~\cite{Non-isolate}.
Moreover, since for large $\epsilon_M$, the qMZMs effectively form a gapped singlet and vanish from the low energy sector, we also restrict ourselves to the most interesting regime $\epsilon_M\ll t_1, t_1'$~\cite{SuppMat}. Our general Hamiltonain then reads $H= H_{\text{DRL}} + H_{\rm Majorana} + H_{\rm M dot}$ where
\begin{align}
\label{eq:HamiltonianFull}
	& H_{\rm Majorana} = i \epsilon_{M}\gamma_1 \gamma_2, \notag \\
	& H_{\rm M dot} = i t_1 \gamma_1 \frac{d_{}^{}+d^{\dag}_{}}{\sqrt{2}} + i t_1' \gamma_1 \frac{d_{}^{}-d^{\dag}_{}}{i\sqrt{2}}\notag \\
	&+ i t_2 \gamma_2 \frac{d_{}^{}+d^{\dag}_{}}{\sqrt{2}}+i t_2' \gamma_2 \frac{d_{}^{}-d^{\dag}_{}}{i\sqrt{2}}.
\end{align}

\textit{Bosonization and basis transformations.--}
The correlator $\tau^{-2r}$ in Eq.~\eqref{eq:dissipation_correlation} suggests that $\varphi$ acts as an effective free bosonic field whose vertex operator $\exp\left(\pm i\varphi\right)$ has the scaling dimension $r$. Dissipation thereby crucially affects the low energy transport in the form of an ``artificial'' Luttinger parameter~\cite{Safi2004, JezouinPierre13, AnthorePRX18}. To reveal this important property, we bosonize the lead operators \cite{GiamarchiBook, LiuRLdissipPRB14}
\begin{align}
	c_{L,R}(x) = F_{L,R}\frac{e^{i\phi_{L,R}(x)}}{\sqrt{2\pi a}},
\end{align} 
where $F_{L,R}$ are Klein factors, $\phi_{L,R}$ are two chiral lead bosons, and $a$ is our UV cufoff (e.g. the lattice constant). We find it further convenient to introduce the following charge and flavor bosons
\begin{align}
	\phi_c= \frac{\phi_L+\phi_R}{\sqrt{2}}, \;\; \phi_f= \frac{\phi_L-\phi_R}{\sqrt{2}}.
\end{align}
With these fields, $\varphi$ and $\phi_f$ always occur with the same sign in Eq.~\eqref{eq:HamiltonianFull}, which suggests the following additional useful linear combinations~\cite{KaneFisherPRB92,LiuRLdissipPRB14}
\begin{align}
	&\phi'_f = \sqrt{\frac{1}{1+r}} \left[\phi_f +\frac{\varphi}{\sqrt{2}} \right],\notag\\
	&\varphi' = \sqrt{\frac{1}{1+r}} \left[\sqrt{r}\phi_f -\frac{\varphi}{\sqrt{2r}} \right].
\end{align}
As a further simplification, we perform the unitary rotation $H\rightarrow UHU^\dagger$, with $U = \exp\left[i(d^\dagger d^{}-1/2)\phi_c(0)/\sqrt{2}\right]$, which removes $\phi_c(0)$ from the exponents in $H_{\text{T}}$. We then obtain the effective tunneling Hamiltonian
\begin{align}
	H_{\rm T} & = \frac{V}{\sqrt{2\pi a}} \Big[F_L d e^{-i\sqrt{1+r}\phi_f'(0)/\sqrt{2}} \notag \\
	& + F_R d e^{i\sqrt{1+r}\phi_f'(0)/\sqrt{2}} + h.c. \Big]\notag \\
	& -i\frac{\pi v_F}{\sqrt{2}}:\psi_c^\dagger(0)\psi_c^{}(0):(d^\dagger d^{}-1/2),
	\label{eq:HTNewNew}
\end{align} 
where $:\psi^{\dagger}_c \psi_c(0): \sim \partial_x \phi_c(0)$ from refermionization. Note that $\varphi'$ completely decouples. In the new basis, the influence of dissipation $r$ on the RG flow of the coupling parameter $V$ is manifest.
Crucially, the last term in Eq.~\eqref{eq:HTNewNew} corresponds to the parallel interaction $J_z$ in the Kondo model~\cite{SchillerIngersent95} and thus modifies the critical temperature $T_c$ below which the system enters the strong coupling regime. However, it has been shown (see e.g. Ref.~\onlinecite{Huaixiu14_1-G}) that its scaling dimension increases upon low energy RG flow and becomes irrelevant. We therefore discard it for our low temperature analysis~\cite{Dressed_parameters}.

\begin{figure}[t!]
\includegraphics[width=0.9 \columnwidth] {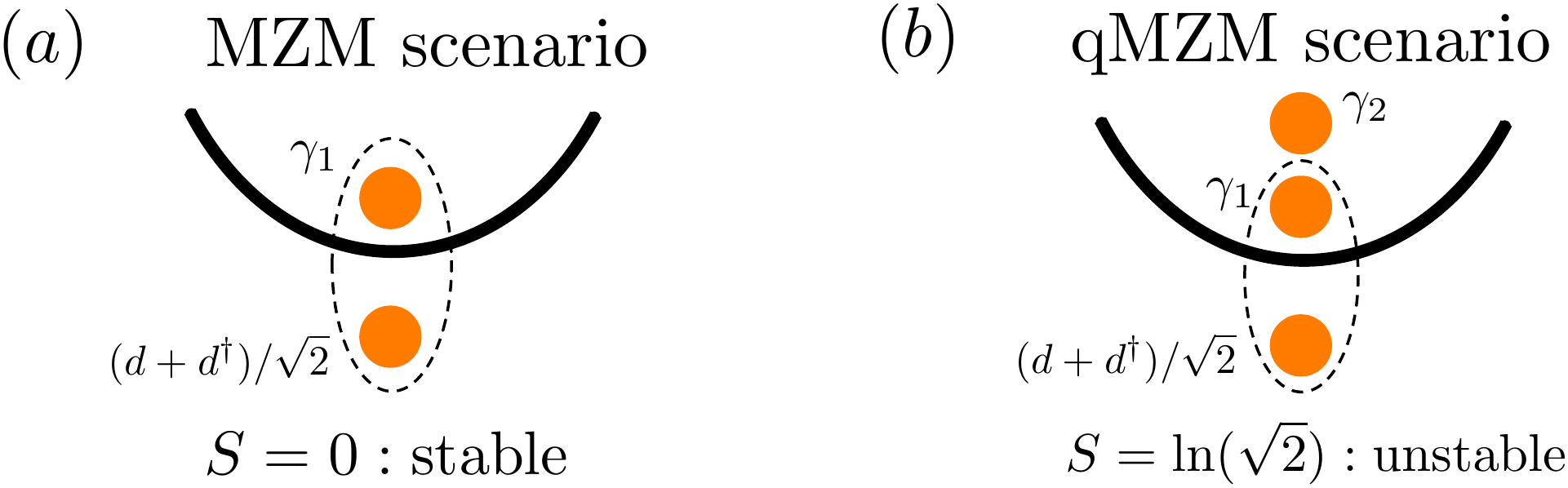}\caption{(a) MZM scenario  and (b) qMZM scenario close to the strong coupling fixed point for $r=1$. The MZM scenario vanishing boundary entropy $S=0$ indicates the stability of the DRL strong coupling fixed point, while the qMZM scenario is unstable and flows instead towards the weak coupling fixed point.}
\label{fig:strong_coupling_impurity}
\end{figure}

\textit{Ground state analysis at the Toulouse point--} The boundary residual entropy of an isolated DRL strong coupling fixed point has been established to be $S=\ln\sqrt{1+r}$~\cite{WongAffleck94}, i.e. it is dissipation dependent. A particularly interesting situation occurs when $r=1$ (the so-called Toulouse point~\cite{EmeryKivelsonPRB92}), where the DRL part of $H$ can be refermionized and solved exactly. Near the strong coupling fixed point, the tunneling Hamiltonian then reads
\begin{align}
	H_{\rm T} & = V [\psi_f^{\dagger}(0) + \psi_f(0) ] (d - d^{\dagger}),
	\label{eq:HTMajorana}
\end{align} 
where $\psi_f \equiv \exp(i\phi_f)/\sqrt{2\pi a}$. From Eq.~\eqref{eq:HTMajorana}, it is clear that while one impurity Majorana operator is involved in the lead-dot coupling, the other decouples and the ground state impurity entropy is therefore non-Fermi liquid like with $S=\ln \sqrt{2}$. Note that this free MZM is generated by frustration (cf. the two-channel Kondo effect~\cite{SchillerHershToulousePRB98}) rather than topology. The other zero-mode partner is effectively located at infinity (since we assume infinite leads) and thus does not enter our analysis.

Coupling the $r=1$ DRL to a topological MZM, it follows that at the strong coupling fixed point, the two available MZMs tend to hybridize and form a singlet, i.e. a Fermi liquid like ground state with $S=0$ (see Fig.~\ref{fig:strong_coupling_impurity}\textcolor{blue}{a}). At the weak coupling fixed point, one MZM instead remains decoupled and again $S=\ln \sqrt{2}$. From the $g$-theorem (see e.g. \cite{AffleckGTheoremPRL91}), which states that for a boundary system, the RG flow is towards the fixed point with the lower boundary entropy (with the bulk staying in the same universality class), it follows that if coupled to a topological MZM, the strong coupling fixed point with $G=e^2/h$ is preferred. Importantly, the direction of the flow is stable against finite $\epsilon_d$, since the MZM provides the necessary degeneracy, otherwise required by fine tuning.

The situation is the very opposite for qMZMs. At low temperatures, the strong coupling fixed point causes the impurity MZM to gap out one of the qMZMs while the other remains, generating $S=\ln \sqrt{2}$ (see Fig.~\ref{fig:strong_coupling_impurity}\textcolor{blue}{b}). At the weak coupling fixed point, all Majorana operators combine into pairs and $S=0$. Hence, the weak coupling fixed point with $G=0$ is preferred. We may therefore conclude that the MZM and qMZM scenarios in the wire cause DRL RG flows towards distinct fixed points with different entropies. These entropies might be detectable by the schemes proposed in several recent studies~\cite{Smirnov2015,Kleeorin2019,Sela2019}.

\begin{figure}[t!]
\includegraphics[width=1.0 \columnwidth] {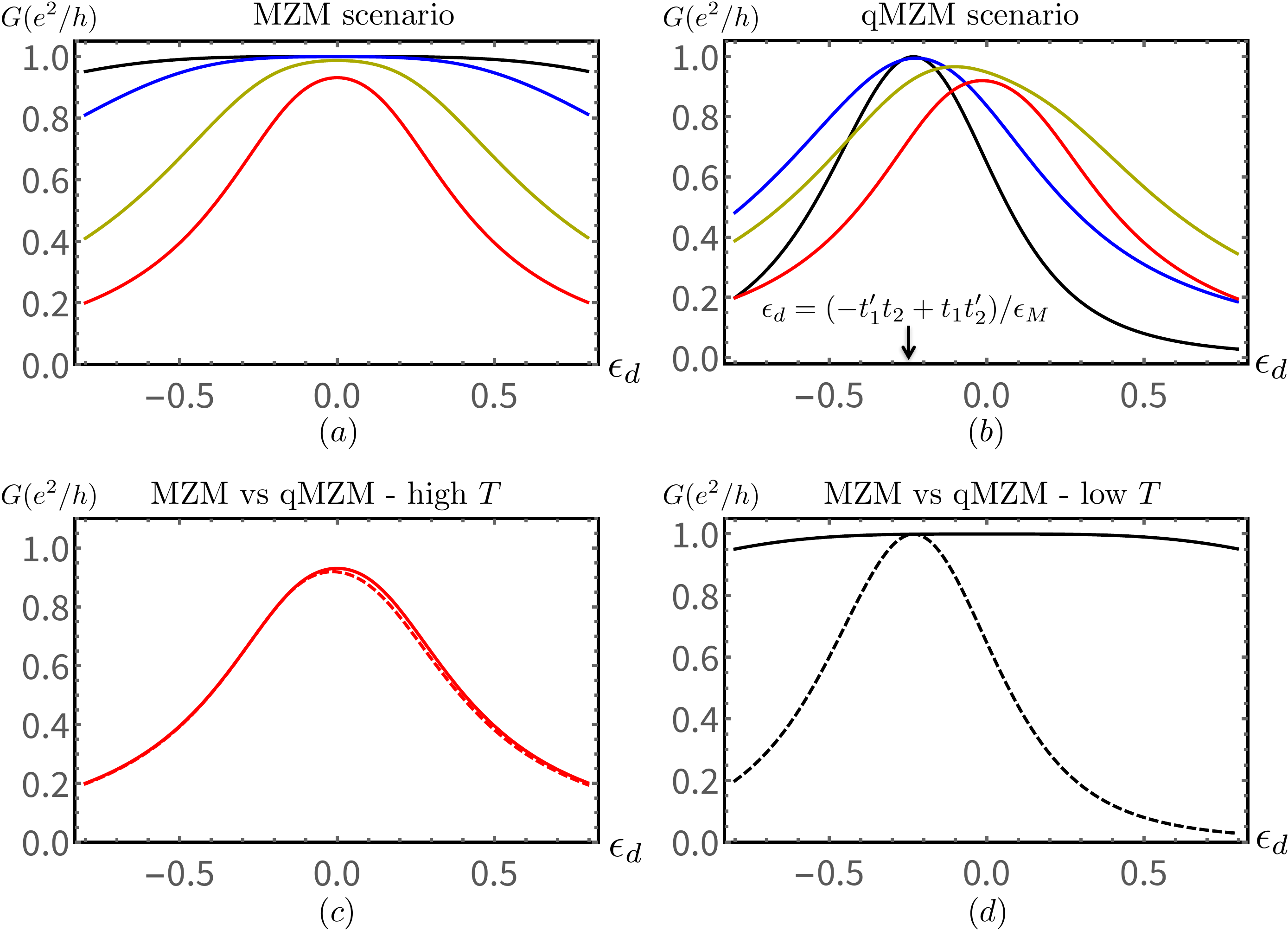}\caption{Dissipative resonant level zero bias conductance $G$ (in units of $e^2/h$) as a function of the detuning $\epsilon_d$ (in units of the dot level broadening $\Gamma$) for various parameter choices. In all figures, $t_1 = -\Gamma$, $t_{1}' = 1.5 \Gamma$, and distinct colors corresponding to different temperatures $T$: Black: $T = 0.002 \Gamma$; Blue: $T = 0.005 \Gamma$; Yellow: $T = 0.02 \Gamma$ and Red: $T = 0.05 \Gamma$. (a)  When the DRL couples to single MZM,  $t_2=t_2'=0$, the peak width increases with decreasing temperature. (b) When coupling the DRL to a pair of qMZMs, with $t_2 = 0.01\Gamma$, $t_2' = 0.02\Gamma$ for the weakly-coupled qMZM and $\epsilon_M = 0.015\Gamma$,  the influence of the weakly coupled qMZM to the conductance curve becomes manifest when $T \sim t_2'$. (c) and (d) Direct comparison of $G$ when coupling to a MZM (solid lines) or two qMZMs (dashed lines) for (c) $T = 0.05\Gamma$ and (d) $T = 0.002 \Gamma$.}
\label{fig:conductance_curves}
\end{figure}

\textit{Conductance scaling.--}
As a probe of the fixed points, we compute the zero bias (differential) conductance through the dot with
\begin{align}
	G = \frac{e^2}{h}\int_{-\infty}^{\infty}d\omega\; C(\omega)[-\partial_\omega n_F(\omega,T)],
	\label{eq:conductance_expression}
\end{align}
where $\omega$ is the energy,  $C(\omega) = A(\omega) \Gamma/2$ is proportional to the dot spectral function $A(\omega)$ obtainable for instance by the equation of motion method \cite{Bruus-Flensberg}, $\Gamma = \pi \rho V^2$ is the dot level broadening, $\rho$ is the lead density of states, and $n_F(\omega,T)$ is the Fermi distribution function. To leading order in $\omega$, we find
\begin{align}
&	C(\omega)\approx 1-\frac{(\epsilon_d^2+t_1^2+t_1'^2)^2\omega^2}{\Gamma^2 t_1^4}, \quad \text{for a MZM,}\label{eq:GMZM}\\
&	C(\omega)\approx \frac{\Gamma^2(\epsilon_M^2+t_1^2+t_2^2)^2\omega^2}{(\epsilon_d \epsilon_M + t'_1 t_2^{} -t_1^{} t_2' )^4}, \quad \text{   for qMZMs.} \label{eq:GqMZMs}
\end{align}

The exact low-temperature zero-bias conductance [i.e.  Eq.~\eqref{eq:conductance_expression}] is presented for various parameters in Fig.~\ref{fig:conductance_curves}.
We see that in the topological MZM scenario, $G=e^2/h$ for the low temperature conductance peaks, which decreases only slightly away from $\epsilon_d=0$. This feature suggests a stabilization of the critical point by the topology-induced degeneracy, i.e., that in the topological Kondo effect. Furthermore, Eq.~\eqref{eq:GMZM} indicates that the peak width $\sim \max(t_1, t_1') \cdot \sqrt{\Gamma/T}$ increases with decreasing temperature, with an onset of enhancement at $T\sim t_1$. At $T=0$, $G=e^2/h$ for all $\epsilon_d$. In contrast, the qMZM scenario presents two other distinct ZBCP features: a shift of the peak position and a vanishing width with decreasing temperature. In particular, at $T=0$, the ZBCP vanishes for all $\epsilon_d$ except when $\epsilon_d \epsilon_M + t'_1 t_2 -t_1 t_2'=0$, where only one qMZM effectively couples to the dot~\cite{SuppMat}.
We therefore conclude that coupling the DRL to qMZMs, the ZBCP shifts from $\epsilon_d=0$ to $\epsilon_d = (-t'_1 t_2 +t_1 t_2')/\epsilon_M$, which signifies the presence of qMZMs. We emphasize that while a small shift is a generic feature of qMZMs with weak $\epsilon_M$, it is here strongly amplified by the dissipation. Furthermore, at finite temperature, a scaling analysis yields a low temperature peak width $\propto T^{1/2}$, identical to that of the isolated DRL model with $r = 1$ \cite{Mebrahtu12,Mebrahtu13}.
We emphasize that Eqs.~\eqref{eq:GMZM} and~\eqref{eq:GqMZMs} refer to distinct fixed points and the former cannot be obtained from the latter by directly setting $t_2=t_2'=\epsilon_d=0$. Indeed, Eq.~\eqref{eq:GqMZMs} holds only for $\omega \sim T < T_2$ (see Fig.~\ref{fig:Scalings}). With decreasing $T_2$, we expect a crossover from the qMZM to the MZM scenario.

\textit{Discussion.--}
While quantum dots have been proposed before in the detection of MZMs (see e.g. Ref.~\cite{Liu2011}), we emphasize here the importance of strong ($r\approx 1\Leftrightarrow \mathcal{R}\approx$ 26k$\Omega$) dissipation. With negligible dissipation, the boundary quantum phase transition, which is highly sensitive towards parameter detuning, disappears.
The zero temperature conductance is then always Lorentzian and its response to temperature variations is comparatively much weaker~\cite{SuppMat}. Dissipation thereby improves the detection of qMZMs by effectively amplifying the weak $\epsilon_M$ and $t_2, t_2'$.

We now argue that relaxing the assumption $r=1$ does not qualitatively change our results as long as $0<r < 2$ \cite{LiuRLdissipPRB14}. The presence of a MZM only gaps out one impurity MZM, leaving the other impurity MZM to bridge the transport through the quantum dot. In comparison, both impurity MZMs are gapped out by two qMZMs, thus blocking the transport at zero temperature. A more detailed RG study~\cite{SuppMat} indicates that near the strong coupling fixed point, the leading transport-reducing operators have scaling dimensions $1 + 1/(1 + r)$ and $1/(1 + r)$ for the MZM and qMZMscenarios, respectively. In turn, these scaling dimensions generate respective width scalings $\sim T^{-1/(1+r)}$ and $T^{r/(1+r)}$ (see e.g. Ref.~\onlinecite{KaneFisherPRB92}). These power-laws reduce to $T^{-1/2}$ and $T^{1/2}$ when $r = 1$, in perfect agreement with Eqs.~\eqref{eq:conductance_expression}-\eqref{eq:GqMZMs}.

We have generally assumed a negligible overlap to the MZM residing on the opposite edge of the wire. This is ensured by using a sufficiently long wire~\cite{Kitaev2001}. It was also recently shown~\cite{Zhang2019} that an additional DRL can be exploited to mitigate such an overlap. Incorporating a small but finite overlap in the MZM scenario corresponds in our model to $\epsilon_M\neq t_2=t_2'=0$ which results in a peak centered at $\epsilon_d=0$ whose width decreases with decreasing $T$~ \cite{SuppMat}. Our model thereby generates unique features for also this scenario.

To roughly estimate the experimental feasibility of our proposal, we consider the setups in Refs.~\onlinecite{Mebrahtu12,Mebrahtu13} for the DRL model and Ref.~\onlinecite{HaoZhangNature18} for the Majorana coupling parameters.
For the DRL system, a dissipation of $r = 0.75$ yielded $G\approx e^2/h$ around $50$mK, which is also approximately the temperature around which the latter reference shows an onset of the low energy state induced ZBCP. 
Under these conditions, assuming further $t_2/t_1\sim 10^{-2}$, a temperature range $\sim 1$ mK should be sufficient to detect the DRL ZBCP width scaling as $\propto T^{-1/(1 + r)}$, or a peak position shift with decreasing temperature. Assuming instead $t_2/t_1\sim 10^{-1}$ (i.e. a sharper wire confinement potential) experimental features should be visible as high as $T\sim 10$ mK. 

Before concluding, we highlight the intricate relation between symmetry and topology. Normally, symmetries of a system provide the possibility of a topologically non-trivial phase (see e.g. Ref.~\onlinecite{ChiuRMP2016} for a review). Our findings here provide an example where \textit{topology replaces a fine-tuned particle-hole symmetry}, and thus serves as a tool to realize phenomena where fine-tuning would otherwise be required.

\textit{Summary.--}
We showed that a DRL coupled to a topological MZM exhibits at low temperatures a ZBCP of height $G=e^2/h$ and a width scaling as $\sim T^{-1/(1 + r)}$.  In contrast, coupling the DRL to a pair of qMZMs, the ZBCP shifts away from resonance and its width shrinks as $\sim T^{r/(1 + r)}$ for $T<T_2$ (see Fig.~\ref{fig:Scalings}). These disparate results follow from flow towards different RG fixed points, where the DRL strong coupling fixed point is guaranteed by the topologically induced degeneracy from a topological MZM. Our findings thereby suggest that \textit{a topological phase transition in the SC nanowire triggers in turn a boundary phase transition} in the DRL, which is detectable in the zero bias conductance. Our proposal can naturally be extended to detect MZMs in other hybrid systems. For example, we envision that one could distinguish between topologically pinned MZMs and trivial low energy states in Josephson junctions (see e.g. Ref.~\onlinecite{Spanslatt2015PiJunc}).

In addition to demonstrating a delicate interplay between MZMs
from topology and from frustration/interactions, we expect
our results to be useful for identifying true MZMs and topological superconductivity, thereby paving the way towards
topologically protected qubits.

\begin{acknowledgments}
\textit{Acknowledgments.--}
GZ thanks Ruixing Zhang for inspiring discussions stimulating this work. We also thank I.V. Gornyi, H.U. Baranger, and R. Willa for helpful comments. This work was supported by the FLAGERA JTC Project GRANSPORT through DFG Grant No. GO 1405/5 and DFG Grant No. MI 658/10-1.
\end{acknowledgments}


\begin{thebibliography}{69}
\bibitem [{2CK()}]{2CK}%
  \BibitemOpen
  \href@noop {} {}\bibinfo {note} {In the DRL model, this degeneracy is
  satisfied through tuning the dot to the particle-hole symmetric point
  $\epsilon_d = 0$. In contrast, the degeneracy occurs naturally in the 2CK
  model from spin SU(2) symmetry.}\BibitemShut {Stop}%
\bibitem [{Non()}]{Non-isolate}%
  \BibitemOpen
  \href@noop {} {}\bibinfo {note} {The value of at least one parameter of
  $t_2$, $t_2'$ and $\epsilon_M$ is assumed to be non-zero. Otherwise
  $\gamma_2$ becomes entirely isolated and thus undetectable with the
  dot.}\BibitemShut {Stop}%
\bibitem [{Dre()}]{Dressed_parameters}%
  \BibitemOpen
  \href@noop {} {}\bibinfo {note} {Note that, by dropping the quartic term we
  ignore its effect in all regimes. All parameters after dropping this term
  thus become effective. We therefore can only determine $T_c$ with the
  complete Hamiltonian Eq.\,\eqref{eq:HTNewNew}. Followng
  Ref.\,\cite{SchillerIngersent95}, we conjecture that $T_c \propto \exp[-
  (\sqrt{2} -\sqrt{(1 + r)/2})^{-1} v_F/V]$.}\BibitemShut {Stop}%
\end{thebibliography}
%

\clearpage
\newpage

\onecolumngrid
\setcounter{equation}{0}
\setcounter{figure}{0}
\setcounter{table}{0}
\setcounter{page}{1}
\renewcommand{\theequation}{S\arabic{equation}}
\renewcommand{\thefigure}{S\arabic{figure}}
\renewcommand{\bibnumfmt}[1]{[S#1]}
\global\long\def\thesection{S\Alph{section}}
\global\long\def\thesubsection{\Roman{subsection}}

\bigskip
\begin{center}
\large{\bf Supplemental Material for ``Distinguishing between Topological and Quasi Majorana Zero Modes with a Dissipative Resonant Level''}
\end{center}
\begin{center}
Gu Zhang$\textcolor{blue}{^{*}}$ and Christian Sp\r{a}nsl\"{a}tt$\textcolor{blue}{^{\dagger}}$\\
{\it Institute for Quantum Materials and Technologies, 76021 Karlsruhe, Germany \\Institut f\"{u}r Theorie der Kondensierten Materie, Karlsruhe Institute of Technology, 76128 Karlsruhe, Germany}\\
(Dated: \today)
\end{center}
In this Supplemental Material, we provide details on the Majorana-dot couplings (Sec.~\ref{sec:MDot}), the shift of the conductance peak in the qMZM scenario (Sec.~\ref{sec:qMZMShift}), and brief RG calculations for the stability of the topologically protected ZBCP with general $0<r<2$ (Sec.~\ref{sec:GeneralRG}). For the special value $r = 1$, we also supply additional conductance curves for other parameters than in the main text (Sec.~\ref{sec:qMZMCurves}) We end with a direct comparison between the DRL and a dissipation free resonant level (Sec.~\ref{sec:DissFree}).
\section{Majorana-dot couplings}
\label{sec:MDot}
In this section, we briefly discuss the parameters $t_1$, $t_1'$, $t_2$ and $t_2'$ of Eq.~\eqref{eq:HamiltonianFull} in the main text. Let us consider a superconducting wire coupled to a dot. This setup is described by electron creation operators $d_1^{\dagger}$ and $d_2^{\dagger}$. In general, the coupling Hamiltonian reads
\begin{equation}
H_{\text{D-dots}} = \lambda_1 d^{\dagger}_1 d_2 + \lambda_1^* d_2^{\dagger} d_1 + \lambda_2 d^{\dagger}_1 d_2^{\dagger} + \lambda_2^* d_2 d_1,
\label{eq:double_dot}
\end{equation}
where the $\lambda_i$ are general coupling parameters. Note that the pair-creation and annihilation operators are to be accompanied by an operator creating or destroying a Cooper pair in the superconductor. However, if the superconductor is large and grounded, and if the fluctuations in expectation value of the electron number operator are negligible, the effects of the pair-operator can be treated as simple phase factors absorbed into $d_{1,2}$ and $d_{1,2}^{\dagger}$: the anti-commutation relation between these operators do not change. We next rewrite $d_1$ and $d_2$ into two pairs of Majorana operators
\begin{equation}
\begin{aligned}
& d_1^{\dagger} = \frac{1}{\sqrt{2}} ( \chi_1 + i \eta_1 ),\ \ \ \  d_1 = \frac{1}{\sqrt{2}} ( \chi_1 -  i \eta_1 ), \\
& d_2^{\dagger} = \frac{1}{\sqrt{2}} ( \chi_2 + i \eta_2 ), \ \ \ \  d_2 = \frac{1}{\sqrt{2}} ( \chi_2 -  i \eta_2 ),
\end{aligned}
\end{equation}
and Eq.\,(\ref{eq:double_dot}) becomes
\begin{equation}
\begin{aligned}
H_{\text{D-dots}} & = i[\text{Im} (\lambda_1) + \text{Im} (\lambda_2)] \chi_1 \chi_2 + i[-\text{Re} (\lambda_1) + \text{Re} (\lambda_2)] \chi_1 \eta_2 \\
&+ i[-\text{Re} (\lambda_1) - \text{Re} (\lambda_2)] \chi_2 \eta_1 + i[\text{Im} (\lambda_1) - \text{Im} (\lambda_2)] \eta_1 \eta_2.
\end{aligned}
\label{eq:hddots}
\end{equation}
Eq.~\eqref{eq:hddots} thus gives us the choice of parameters used in Eq.~\eqref{eq:HamiltonianFull} in the main text. Without loss of generality, we can always get rid of the phases in $\lambda_{1,2}$ by a redefinition of the dot operators such that both $\lambda_1$ and $\lambda_2$ become real. Then only $\chi_1 \eta_2$ and $\chi_2 \eta_1$ remain and the situation corresponds to the $t_1, t_2' = 0$ case  in Eq.~\eqref{eq:HamiltonianFull}. However, our definition of parameters in Eq.~\eqref{eq:HamiltonianFull} is not redundant: there, the $d$ and $d^{\dagger}$ operators are defined such that $V$ is real, and we do not have any additional choice to re-define them. This leaves us with the general form of $H_{\rm Mdot}$ in Eq.~\eqref{eq:HamiltonianFull}.

\section{Shift of the conductance peak for qMZMs}
\label{sec:qMZMShift}
Here, we comment upon the shift of the conductance peak from $\epsilon = 0$ to $\epsilon_d = (- t_1' t_2 + t_1 t_2')/\epsilon_M$  in the qMZM scenario. To understand this, we consider $V=0$ in Eq.~\eqref{eq:ht} in the main text. The system then consists of four Majorana fermions coupling to each other and can be straightforwardly diagonalized. The low energy states have energies
\begin{equation}
\epsilon_{\pm} = \frac{1}{2} [ \epsilon_d - \sqrt{\epsilon_d^2 + t_1^2 + t_1'^2 + t_2^2 + t_2'^2 \pm 2( \epsilon_d \epsilon_M + t_1' t_2 - t_1 t_2')} ].
\label{eq:gs_energy}
\end{equation}
Apparently, when $\epsilon_d \epsilon_M + t_1' t_2 - t_1 t_2' = 0$ we have that  $\epsilon_+ = \epsilon_-$ and the impurity system consists of two degenerate ground states. At this very point, the system thus contains a double degeneracy even in the qMZM scenario. For low temperatures, the system then mimics that of a bare DRL model except that the degenerate symmetric point now occurs at $\epsilon_d = (- t_1' t_2 + t_1 t_2')/\epsilon_M$ instead of $\epsilon_d = 0$.

\section{General RG calculations}
\label{sec:GeneralRG}
For the special choice $r=1$, we discussed in the main text the RG flow of the system by using the $g$-theorem. For a general choice of $r$, we investigate here the flow with RG arguments. For convenience, we first define the two impurity Majorana fermions $a = (d + d^{\dagger})/\sqrt{2}$ and $b = (d - d^{\dagger})/(i\sqrt{2})$.

\subsection{RG equations in the MZM scenario}
When the wire is in the topological regime, only one Majorana operator is present in Eq.~\eqref{eq:HamiltonianFull} in the main text. In what follows, we denote it simply by $\gamma$. The effective tunneling Hamiltonian and the dot-Majorana coupling then read
\begin{equation}
\begin{aligned}
H_{\text{T}} + H_{\text{dot}} + H_{\text{Mdot}} & = iV \sqrt{\frac{2}{\pi a}} F\cos[\sqrt{\frac{1+r}{2}} \phi_f'(0)]b + i\epsilon_d a b + i t_1 a \gamma + i t_1' b \gamma,
\end{aligned}
\label{eq:real_mzm_effective_ham}
\end{equation}
where we have already ignored the quartic term [the last term of Eq.~\eqref{eq:HTNewNew}]. In this expression, the influence of $t_1'$ is quite limited since it can be combined with the dot detuning into an effective detuning $i(\epsilon_d a - t_1' \gamma) b$. More interesting is the influence of $t_1$.

To begin with, if $t_1 = 0$, the effective Hamiltonian becomes that of a detuned isolated DRL. At low temperatures, the leading operator that may destroy the conductance peak, i.e. the effective detuning operator, becomes~\cite{GUSM}
\begin{equation}
\sim \tilde{t} \sin[\sqrt{\frac{2}{1+r}} \phi_f'] (\epsilon_d a - t_1' \gamma),
\label{eq:effective_low_temp_topo}
\end{equation}
where $\tilde{t}$ is the weak backscattering parameter dual to $\tilde{V}$. The leading order RG equation for $\tilde{t}$ reads
\begin{align}
\frac{d\tilde{t}}{d\ell} = \left(1-\frac{1}{1+r}\right)\tilde{t},
\end{align}
i.e. the scaling dimension of $\tilde{t}$ is $1/(1+r)$, and $\tilde{t}$ is relevant for $r > 0$. Hence, any detuning is capable of destroying the resonance feature of an isolated DRL.

On the contrary, if $t_1 \neq 0$, when the temperature $T \ll t_1$, then $a$ and $\gamma$ form a singlet.
Since the impurity part of Eq.~\eqref{eq:effective_low_temp_topo} [i.e, $(\epsilon_d a - t_1' \gamma)$] changes the occupation of this singlet, only virtual states are allowed. Consequently, the leading relevant operator is either the first descendent~\cite{CFTSM} of Eq.~\eqref{eq:effective_low_temp_topo} with scaling dimension $1 + 1/(1 + r)$, or the high-order tunneling parameter
\begin{equation}
\propto \sin[\sqrt{\frac{8}{1+r}} \phi_f'],
\end{equation}
with scaling dimension $4/(1 + r)$. With $0<r<2$, the leading irrelevant operator thus has the scaling dimension $1 + 1/(1 + r) >1$, and the strong coupling fixed point is stable.

\subsection{RG Equations in the qMZM scenario}
In the presence of a pair of quasi-MZMs, the low-temperature system effective Hamiltonian reads
\begin{equation}
H_{\text{T}} + H_{\text{dot}} + H_{\text{Mdot}} = iV \sqrt{\frac{2}{\pi a}} F\cos[\sqrt{\frac{1+r}{2}} \phi_f'(0)]b + i\epsilon_d a b + i t_1 a \gamma_1 + i t_1' b \gamma_1 + i t_2 a \gamma_2 + i t_2' b \gamma_2.
\label{eq:quasi_mzm_ham}
\end{equation}
Once again, the terms $i t_1' b \gamma_1$ and $i t_2' b \gamma_2$ are not important since they combine with the term $i\epsilon_d a b$ into $i(\epsilon_d a - t_1' \gamma_1 - t_2' \gamma_2) b$ into an effective dot detuning. We now study the effects of the terms $i t_1 a \gamma_1 + i t_2 a \gamma_2 + i \cdot 0 \cdot a \zeta$, where $\zeta$ is an auxiliary Majorana operator introduced for the convenience of the analysis.

We consider a system that consists of two two-level systems, and label their states as $|n_1, n_2 \rangle$. Here, $n_1=0,1$ labels the fermionic state composed by $a$ and $\zeta$, and $n_2=0,1$ that of $\gamma_1$ and $\gamma_2$. We next diagonalize this composite system to obtain the eigenstates. Interestingly, the ground state with energy $-\sqrt{t_1^2 + t_2^2}/2$, is doubly degenerate with
\begin{equation}
|g_1\rangle = \frac{1}{\sqrt{2}}\Big[\frac{i t_1 + t_2}{\sqrt{t_1^2 + t_2^2}} |0,0\rangle + |1,1\rangle \Big],\ \ \text{and}\ \ |g_2\rangle = \frac{1}{\sqrt{2}}\Big[\frac{-i t_1 - t_2}{\sqrt{t_1^2 + t_2^2}} |1,0\rangle + |0,1\rangle \Big].
\label{eq:ground_states}
\end{equation}

From Eq.~\eqref{eq:ground_states} we see that in comparison to the MZM scenario, the system now has the possibility to transition between $|g_1\rangle$ and $|g_2\rangle$. The effective backscattering of the DRL at low temperature then becomes
\begin{equation}
\sim \tilde{t} \sin\Big[ \sqrt{\frac{2}{1 + r}} \phi_f' \Big] (c_1\sigma_x + c_2 \sigma_y),
\end{equation}
where $\sigma_x$ and $\sigma_y$ are two Pauli matrices in the space spanned by $|g_1\rangle$ and $|g_2\rangle$, and $c_1$ and $c_2$ are two non-universal constants. The backscattering operator has the scaling dimension $1/(1+r)$ and becomes relevant when $r>0$. Consequently, the weak backscattering at the strong coupling fixed point becomes relevant, and the system prefers the weak coupling fixed point.

\section{Additional conductance curves for $r=1$}
\label{sec:qMZMCurves}

In contrast to the parameter choices of Fig.~\ref{fig:conductance_curves} in the main text, 
we provide here conductance curves in two additional regimes.
Experimentally, one may reach these regimes for sharper wire confinement potentials.

\begin{figure}[]
\includegraphics[width=0.8\columnwidth] {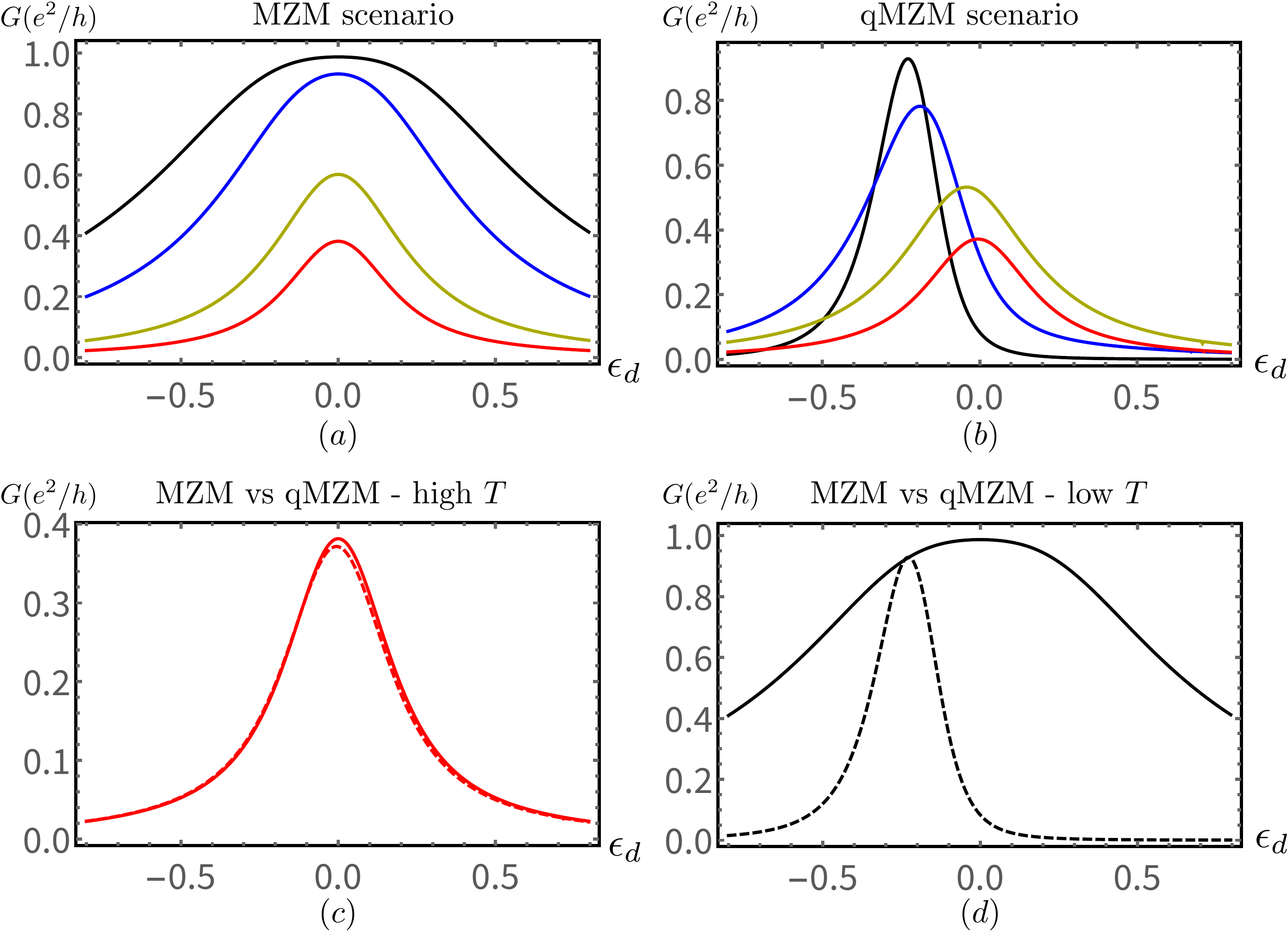}\caption{Dissipative resonant level zero bias conductance $G$ (in units of $e^2/h$) as a function of the detuning $\epsilon_d$ (in units of the dot level broadening $\Gamma$) under different temperatures. The temperature $T = 0.5\Gamma, 0.2\Gamma, 0.05\Gamma$ and $0.02\Gamma$ for the red, yellow, blue, and black curves respectively.  Here we take $t_2 = 0.1\Gamma$ and $t_2' = 0.2\Gamma$. Other parameter choices are the same in making Fig.\,\ref{fig:conductance_curves} of the main text.}
\label{fig:conductance3}
\end{figure}

As our first regime, we take $t_2 /t_1 \sim 10^{-1}$, which generate the conductance curves depicted in Fig.~\ref{fig:conductance3}. We notice that a manifest peak shift now begins to emerge at a much higher temperature $T \sim 0.1 \Gamma$ than in Fig.~\ref{fig:conductance_curves} of the main text. Experimentally, this ratio allows for a detection of qMZMs at $T\approx 10$ mK.

\begin{figure}[t!]
\includegraphics[width=0.8\columnwidth] {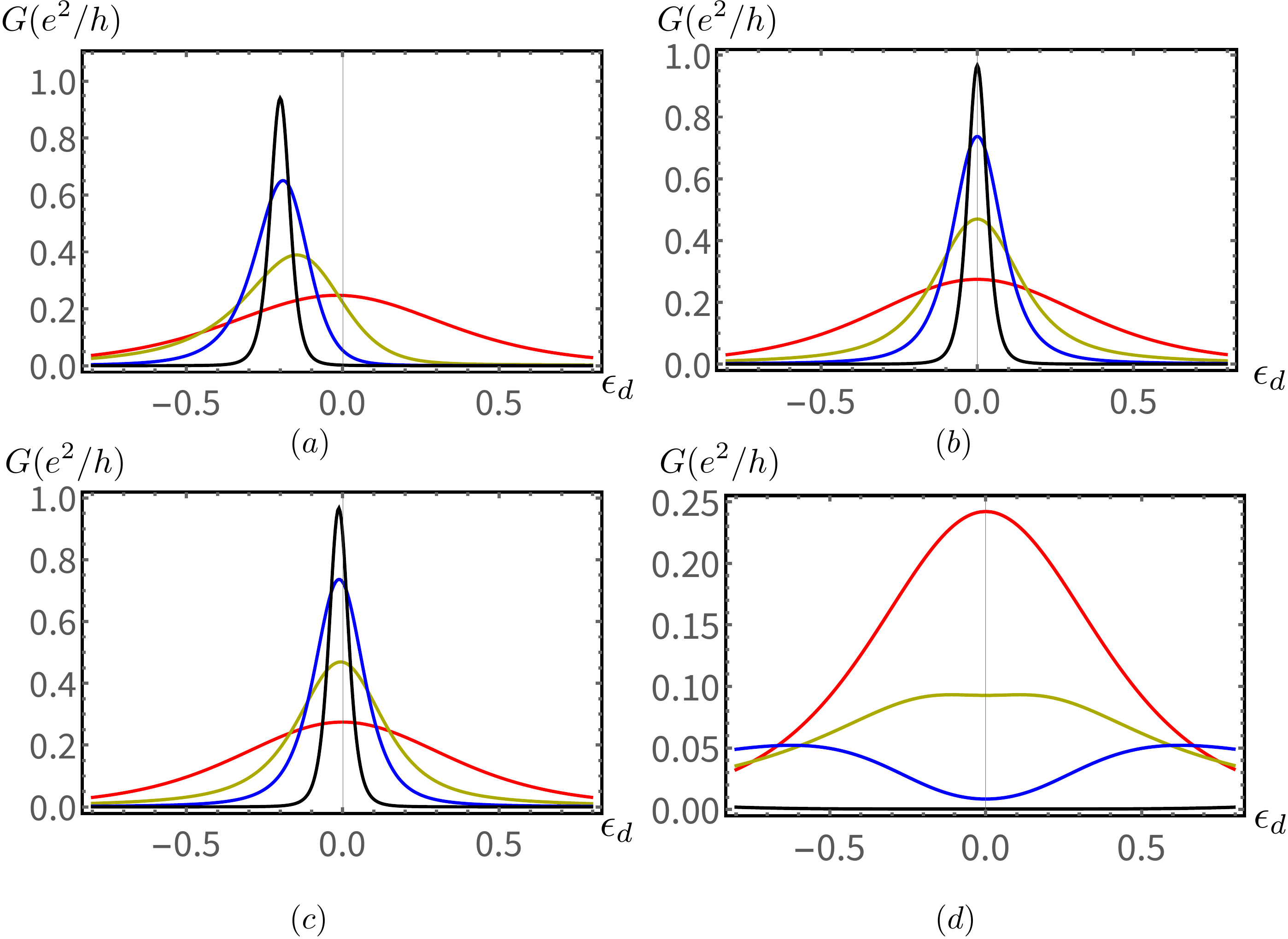}\caption{
Dissipative resonant level zero bias conductance $G$ (in units of $e^2/h$) as a function of the detuning $\epsilon_d$ (in units of the dot level broadening $\Gamma$) for various parameter choices of $t_2$, $t_2'$ and $\epsilon_M$. The temperature $T= 2\Gamma, 0.5 \Gamma, 0.2 \Gamma$ and $0.05 \Gamma$ for the red, yellow, blue, and black curves respectively. These temperatures are larger than those in the main text.
The values of $t_1$ and $t_1'$ are taken as Fig.~\ref{fig:conductance_curves} in the main text.(a) $t_2 = \Gamma$, $t_2' = 1.5\Gamma$ and $\epsilon_M = 1.5 \Gamma$. (b) $t_2 = 0$, $t_2' = 0$ and $\epsilon_M = \Gamma$. (c) $t_2 = 0.05 \Gamma$, $t_2' = 0.05 \Gamma$ and $\epsilon_M = \Gamma$. (d) $t_2 = \Gamma$, $t_2' = 1.5 \Gamma$ and $\epsilon_M = 0$.}
\label{fig:conductance_of_different_cases}
\end{figure}

As our the second regime, we take either $t_2, t_2'$, or $\epsilon_M$ (or all three parameters) to be comparable to $t_1$ and $t_1'$. The results are shown in Fig.~\ref{fig:conductance_of_different_cases}. A comparison with Fig.~\ref{fig:conductance_curves} in the main text indicates that when two qMZMs have similar coupling strengths to the DRL, we can distinguish a topological MZM from qMZMs from the conductance for temperatures as high as $T \sim t_1$.


\begin{figure}[b!]
\includegraphics[width=0.8\columnwidth] {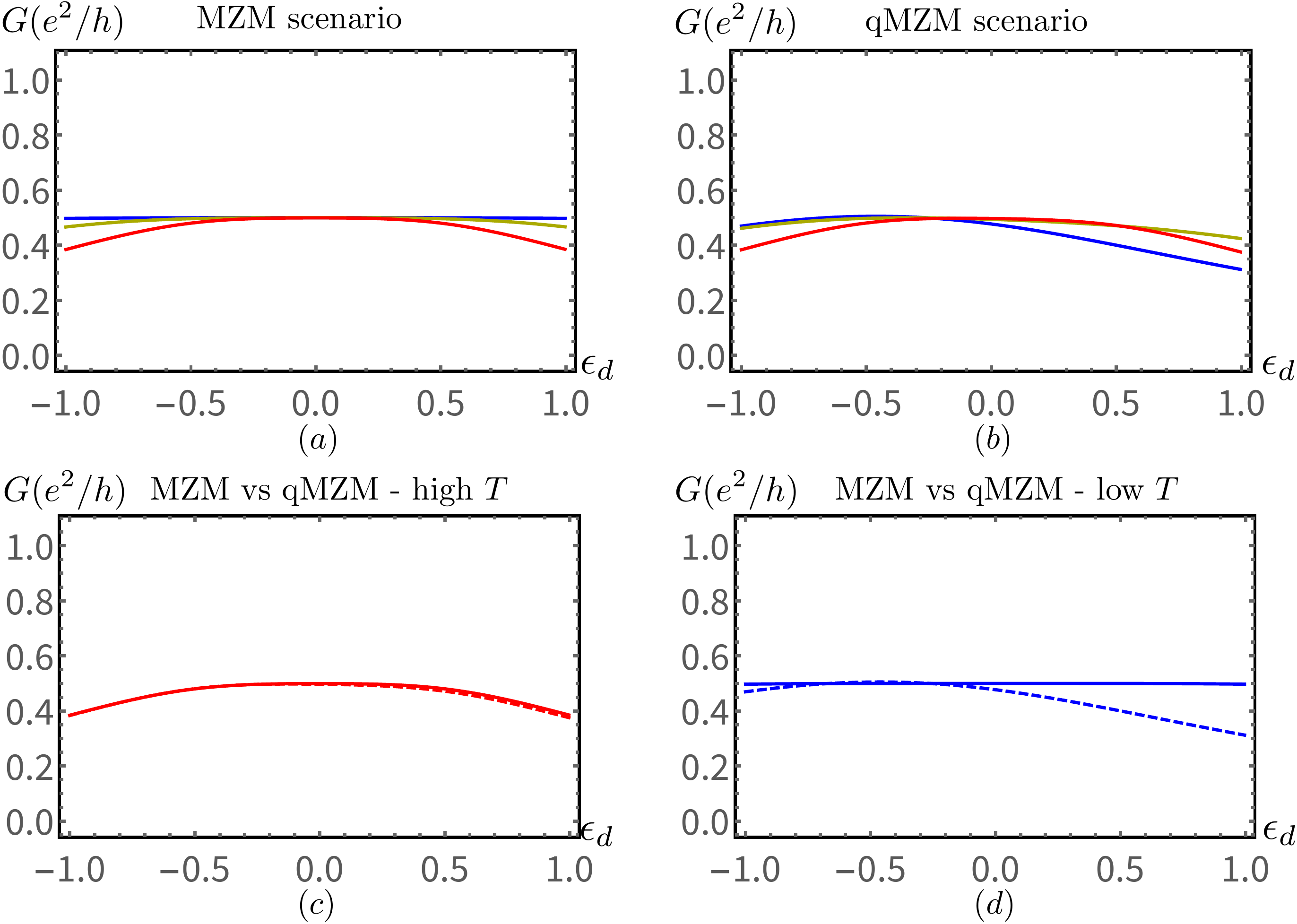}\caption{
Dissipation-free resonant level zero bias conductance $G$ (in units of $e^2/h$) as a function of the detuning $\epsilon_d$ (in units of the dot level broadening $\Gamma$) for different temperatures $T = 0.05 \Gamma$ (red curves), $0.02\Gamma$ (yellow curves) and $0.005\Gamma$ (blue curves).
The other parameters are the same as in Fig.~\ref{fig:conductance_curves} in the main text.
(a) The qMZM scenario. (b) The topological MZM scenario. (c) and (d) direct comparisons between $G$ of the MZM (solid lines) and that of the qMZM (dashed lines) scenarios, under temperatures $T= 0.005\Gamma$ (in blue) and $T = 0.05 \Gamma$ (in red), respectively.
}
\label{fig:dissipation_free}
\end{figure}

\section{Conductance for a Dissipation-free resonant level}
\label{sec:DissFree}
In the main text, we emphasized that the presence of dissipation amplifies the effects of topology: The topologically induced degeneracy stabilizes the quantum critical point of the DRL model, which leads to a strong signature in the presence of the MZM. To illustrate this point further, we provide here conductance curves for a dissipation free resonant level in Fig.~\ref{fig:dissipation_free}. From Fig.~\ref{fig:dissipation_free}(a) which depicts the qMZM scenario, we see that the conductance peak is now Lorentzian, with a width saturated to the level broadening of the resonant level at low temperatures. This feature evidently makes it very difficult to experimentally distinguish between the MZM and qMZM scenarios without the dissipation.

\end{document}